\documentclass[manuscript]{aastex}

\shorttitle{Relativistic Combustion Waves}
\shortauthors{Gao \& Law}

\begin{document}


\title{Rankine-Hugoniot Relations in Relativistic Combustion Waves}


\author{Yang Gao\altaffilmark{1} and Chung K. Law\altaffilmark{1,2}}
\affil{$^1$ Center for Combustion Energy and Department of Thermal Engineering,\\ Tsinghua University,
  Beijing, 100084, China}
\affil{$^2$ Department of Mechanical and Aerospace Engineering,\\ Princeton University,
  Princeton, NJ 08544-5263, USA} \email{cklaw@Princeton.EDU}







\begin{abstract}
As a foundational element describing relativistic reacting waves of relevance to
  astrophysical phenomena, the Rankine-Hugoniot relations classifying the various propagation modes of detonation and deflagration
  are analyzed in the relativistic regime, with the results properly degenerating to the non-relativistic and highly-relativistic limits.
The existence of negative-pressure downstream flows
  is noted for relativistic shocks, which could be of interest in the understanding of the nature of dark energy.
Entropy analysis for relativistic shock waves are also performed for relativistic fluids with different equations of state (EoS),
  denoting the existence of rarefaction shocks in fluids with adiabatic index $\Gamma<1$ in their EoS.
The analysis further shows that weak detonations and strong deflagrations, which are rare phenomena in terrestrial environments,
  are expected to exist more commonly in astrophysical systems because of the various endothermic reactions present therein.
Additional topics of relevance to astrophysical phenomena are also discussed.

\end{abstract}





\section{Introduction}

The classical theory of combustion waves is well established
  in the study of reactive fluid dynamics \citep{williams1985,zeldovich1985,law2006}.
Different from hydrodynamic shock waves which have been extensively studied, supersonic detonations and subsonic deflagrations are
  sustained by reactions in the fronts of fluid discontinuities \citep[see, e.g.,][]{landau1959,williams1985},
  and as such are expected to yield rich varieties of fluid dynamical responses.

Axford and Newman \citep{axford1961,newman1968} first adopted the concept of reactive fluid dynamics in astrophysics by
  considering
  the dynamics of the hydrogen ionization fronts around stars.
\citet{blandford1978,blandford1980} advanced the mechanism of particle acceleration through
  astrophysical shocks such as those in supernova systems, which in principle should also bear the dynamics
  of reactive flows.
Indeed, it is now commonly believed that nuclear deflagration and detonation waves support the explosion of Type Ia supernovae
  \citep{arnett1969,hillebrandt2000,gamezo2003,gamezo2005}, recognizing nevertheless that
  the transition mechanism from deflagration to detonation is still not clear.
Additional studies of astrophysical processes that have suggested/invoked combustion processes include relativistic
  detonation waves as a possible mechanism for the false vacuum decay \citep{steinhardt1982}, relevant for
  the microwave background fluctuations \citep{gibson2005} in the early universe;
the fireball model for the $\gamma$-ray burst \citep{meszaros2002} involving relativistic $\gamma$ photons and
  electron/positron pairs;
and the unique role of the ISM shocks in various chemical processes in molecular clouds, particularly in early-phase star
  formations \citep[e.g.,][]{mckee2007,flower2010} as observed in detail by Herschel and other facilities.
Furthermore, a recent study \citep{gao2011} on the evolution of supernovae remnants has also incorporated reaction
  to account for the accelerative expansion of the Crab Nebula.
It is thus clear that integration of combustion theory offers rich potential in the study of astrophysical phenomena.

To adopt reactive fluid dynamics in the study of astrophysical systems,
  a general form of combustion wave theory in relativistic fluids is needed.
While theories of relativistic shock waves \citep{taub1948} and detonation waves
  in highly-relativistic fluids\footnote{Here highly-relativistic fluids refer to those hydrodynamic systems whose speed of fluid particles are very close, or equal, to the speed of light.}
  \citep{steinhardt1982} have been advanced,
  and the relativistic shock waves for radiating fluids \citep{cissoko1997} and for fluids in magnetic fields \citep{mallick2011}
  have been presented,
  a general analysis of combustion waves describing subsonic deflagration as well as supersonic detonation waves,
  for all relativistic fluids, has not been performed.
Consequently, as a first, necessary step, we shall integrate the essential features of relativistic fluids
  \citep{landau1959,anile1989} and combustion waves \citep{williams1985,law2006}, within the context of the Rankine-Hugoniot relations, and identify the various possible relativistic combustion waves and their properties.
The relativistic theory of deflagration and detonation waves has also been studied in quark-gluon plasmas
  using the bag equation of state for the quark matter \citep{gyulassy1984},
  which is a useful reference for the present study of combustion waves in a Synge gas \citep{synge1957}
  with the adiabatic index $\Gamma$ for more general astrophysical applications.

Since the relativistic shock is an important component in relativistic flows, it will be separately studied first in Section 2,
  in which an interesting solution involving negative pressure in the downstream fluid is identified,
  and entropy analysis to the shocks are founded.
In Section 3, the general relativistic form of deflagration and detonation waves is presented, with its proper degeneracy to the
  non-relativistic
  and highly-relativistic limits.
Detonation and deflagration waves, as well as special cases of Chapman-Jouguet waves and isobaric waves are analyzed therein.
The possible existence of weak detonations and strong deflagrations in relativistic astrophysical environments
  is then discussed, which is followed by summary of the present work, in Section 4.

\section{Relativistic shock waves}

We first briefly outline the fundamentals of relativistic fluid dynamics and the theory of relativistic shock waves
  \citep{taub1948,landau1959,liang1977}.

\subsection{Basic equations governing the relativistic gas}

Based on the energy-momentum tensor in the local rest frame of a relativistic fluid
  \begin{equation}
  T^{ik}=\left(
    \begin{array}{cccc}
    e & 0 & 0 & 0 \\
    0 & p & 0 & 0 \\
    0 & 0 & p & 0 \\
    0 & 0 & 0 & p
    \end{array}\right),
    \label{equ:tensor}
  \end{equation}
  momentum and energy conservations across a shock front are:
  \begin{equation}
  \omega_1 u_1^2 + p_1 = \omega_2 u_2^2 + p_2
  \label{equ:momentum cons}
  \end{equation}
  and
  \begin{equation}
  \omega_1 \gamma_1 u_1 = \omega_2 \gamma_2 u_2,
  \label{equ:energy cons}
  \end{equation}
  respectively \citep{landau1959,steinhardt1982}.
Here $e$ is the fluid energy density, which includes the rest-frame energy $nm_0c^2$ of the fluid particles
  and the specific internal energy $\epsilon$ (see Equation (\ref{equ:internal energy})),
  $p$ is the pressure of the fluid, $\omega=e+p$ the specific enthalpy per unit volume,
  $m_0$ the rest mass of one particle and $n$ the particle number density which can be the number density of any charge (e.g., baryon number or lepton number) that is conserved across the shock front.
The four-velocity $u$ of the fluid has the form $u=\beta \gamma$, where $\beta=v/c$ is the velocity in unit of
  the speed of light and $\gamma=1/(1-\beta^2)^{1/2}$ is the Lorentz factor.
Subscripts 1 and 2 denote upstream and downstream quantities, respectively.
The particle number conservation
  \begin{equation}
  n_1 u_1 = n_2 u_2
  \label{equ:number cons}
  \end{equation}
  is a complementary condition of continuity across the shock.

For analysis pertinent to the stellar interior, the ISM or intergalactic media (IGM), and the early universe,
  the relativistic ideal gas is described by the following equation of state (EoS)
  with the adiabatic index $\Gamma$ \citep[i.e., the Synge gas, cf.][]{synge1957,lanza1982,cissoko1997}:
  \begin{equation}
  \Gamma=1+\frac{p}{\rho\epsilon},
  \label{equ:EoS}
  \end{equation}
  where $\rho=nm_0$ is the rest-frame mass density and $\epsilon$ the specific internal energy
  (cf. Equation(\ref{equ:internal energy})).
Then the fluid energy density can be written as
  \begin{equation}
  e=\rho(c^2+\epsilon)=nm_0c^2+\frac{1}{\Gamma-1}p,
  \label{equ:internal energy}
  \end{equation}
  which can also be considered as the definition of the specific internal energy $\epsilon$.
Using the EoS (\ref{equ:EoS}), the specific enthalpy has the following form \citep{kennel1984},
  \begin{equation}
  \omega=e+p=nm_0c^2+\frac{\Gamma}{\Gamma-1}p.
  \label{equ:enthalpy}
  \end{equation}

The EoS with $4/3\leq\Gamma\leq5/3$ \citep{taub1948,anile1989} is a general form for relativistic fluids.
In the highly-relativistic regime ($\Gamma=4/3$), the internal energy greatly exceeds the rest-frame energy of the particle,
  i.e., $\epsilon\gg c^2$, and the fluid is radiation dominant with $p=\frac{1}{3}\rho \epsilon = \frac{1}{3}e$.
This is the EoS used in \citet{steinhardt1982} in the study of the bubble growth during the false vacuum decay
  in the early evolution phase of the universe.
In the non-relativistic extreme ($\Gamma=5/3$, cf. Anile 1989), $\epsilon\ll c^2$, the EoS assumes the form
  $p=\frac{2}{3}\rho \epsilon$.
Comparing this expression with the classical ideal gas law $p=\rho k T$, with $k$ being the Boltzmann constant,
  the internal energy is just the kinetic energy $\epsilon = \frac{3}{2} k T$ for a monatomic gas
  in non-relativistic fluids.
Another special EoS with $\Gamma=2/3$ accounting for the dark energy in the universe will be discussed in Section 2.3.

The sound speed in relativistic fluids is given by \citep[cf.][]{landau1959}
  \begin{equation}
  v_{\rm s}=c\sqrt{\frac{\partial p}{\partial e}},
  \label{equ:sound speed}
  \end{equation}
  whose four-velocity is $u_{\rm s}=\beta_{\rm s} \gamma_{\rm s}=\frac{v_{\rm s}}{c}(1-(\frac{v_{\rm s}}{c})^2)^{-1/2}$.
Following the above discussion, we have $v_{\rm s}=c/\sqrt{3}$ in the highly-relativistic regime and
  $v_{\rm s}=\sqrt{\frac{\partial p}{\partial \rho}}$ in the non-relativistic regime.

\subsection{The relativistic shock adiabat}

By defining a variable $x=\omega/n^2$ for fluids in both sides of the wave front, we readily
  obtain the expressions for the particle flux ($j=n_1u_1=n_2u_2$)
  by considering momentum conservation (\ref{equ:momentum cons}) and particle flux conservation (\ref{equ:number cons})
  \begin{equation}
  -j^2=\frac{p_2 - p_1}{x_2 - x_1},
  \label{equ:number flux}
  \end{equation}
  and the so-called shock adiabat by additionally considering the energy conservation (\ref{equ:energy cons}) \citep[cf.][]{taub1948,landau1959,steinhardt1982}:
  \begin{equation}
  x_2\omega_2-x_1\omega_1=(p_2-p_1)(x_2+x_1).
  \label{equ:shock adiabatic}
  \end{equation}
Referring to the form of the specific enthalpy (\ref{equ:enthalpy}) under the EoS, the variable $x$ can be expressed as
  \begin{equation}
  x=\frac{1}{n^2}\bigg(nm_0c^2 + \frac{\Gamma}{\Gamma-1}p\bigg)=\frac{m_0^2c^2}{\rho}+
  \frac{\Gamma}{\Gamma-1}\frac{m_0^2 p}{\rho^2}.
  \label{equ:enthalpy 2}
  \end{equation}

Introducing Equation (\ref{equ:enthalpy 2}) to the particle flux (\ref{equ:number flux}), we obtain the
  Rayleigh relation for relativistic fluids, accounting for the conservation of number density and momentum:
  \begin{equation}
  (\hat{p}-1)= -u_1^2 \bigg[\hat{c}^2(\hat{V}-1)+\frac{\Gamma}{\Gamma-1}(\hat{p}\hat{V^2}-1)\bigg].
  \label{equ:Rayleigh 1}
  \end{equation}
Here $V=1/\rho$ is the specific volume, and notations with hats are reduced variables, i.e.,
  $\hat{p}=p_2/p_1$, $\hat{V}=V_2/V_1$ and $\hat{c}=c/\sqrt{p_1V_1}$.
In the limit of $\hat{c}\rightarrow \infty$ and $u_1\rightarrow 0$, the Rayleigh relation degenerates to that in the
  non-relativistic regime \citep[cf.][]{landau1959}:
  \begin{equation}
  M_1^2 = -\frac{\hat{p}-1}{\Gamma(\hat{V}-1)},
  \label{equ:Rayleigh non}
  \end{equation}
  where $M=v/\sqrt{\Gamma p/\rho}$ is the non-relativistic Mach number; while in the limit of $\hat{c}\rightarrow 0$,
  it assumes the highly-relativistic form:
  \begin{equation}
  u_1^2= -\frac{(\Gamma-1)(\hat{p}-1)}{\Gamma(\hat{p}\hat{V}^2-1)}.
  \label{equ:Rayleigh super}
  \end{equation}

By introducing $x$ given by (\ref{equ:enthalpy 2}) to the shock adiabat (\ref{equ:shock adiabatic}),
  we derive the Hugoniot relation for relativistic shocks:
  \begin{equation}
  \hat{c}^2\bigg[\frac{\Gamma+1}{\Gamma-1}(\hat{p}\hat{V}-1)-(\hat{p}-\hat{V})\bigg]
  =\frac{\Gamma}{\Gamma-1}\hat{p}(1-\hat{V}^2) - \frac{\Gamma}{(\Gamma-1)^2}(\hat{p}^2\hat{V}^2-1).
  \label{equ:Hugoniot 1}
  \end{equation}
The Hugoniot relation additionally takes into account of energy conservation (\ref{equ:energy cons})
  and as such describes all possible discontinuities across a shock by considering
  all conservation equations.
In the non-relativistic extreme of $\hat{c}\rightarrow \infty$ and $u_1\rightarrow 0$,
  Equation (\ref{equ:Hugoniot 1}) reduces to \citep{landau1959}
  \begin{equation}
  \bigg(\hat{p}+\frac{\Gamma-1}{\Gamma+1}\bigg)\bigg(\hat{V}-\frac{\Gamma-1}{\Gamma+1}\bigg)
  = \frac{4\Gamma}{(\Gamma+1)^2}.
  \label{equ:hugoniot non}
  \end{equation}
In the highly-relativistic extreme of $\hat{c}\rightarrow 0$, the RHS of Equation (\ref{equ:Hugoniot 1}) vanishes, yielding
  \begin{equation}
  \hat{p}\hat{V}^2=\frac{(\Gamma-1)\hat{p}+1}{(\Gamma-1)+\hat{p}}.
  \label{equ:Hugoniot super}
  \end{equation}
The highly-relativistic shock adiabat, i.e., Rayleigh and Hugoniot relations, have been derived and discussed in, e.g.,
  \citet{steinhardt1982}.

\begin{figure}
 \epsscale{1.0} \plotone{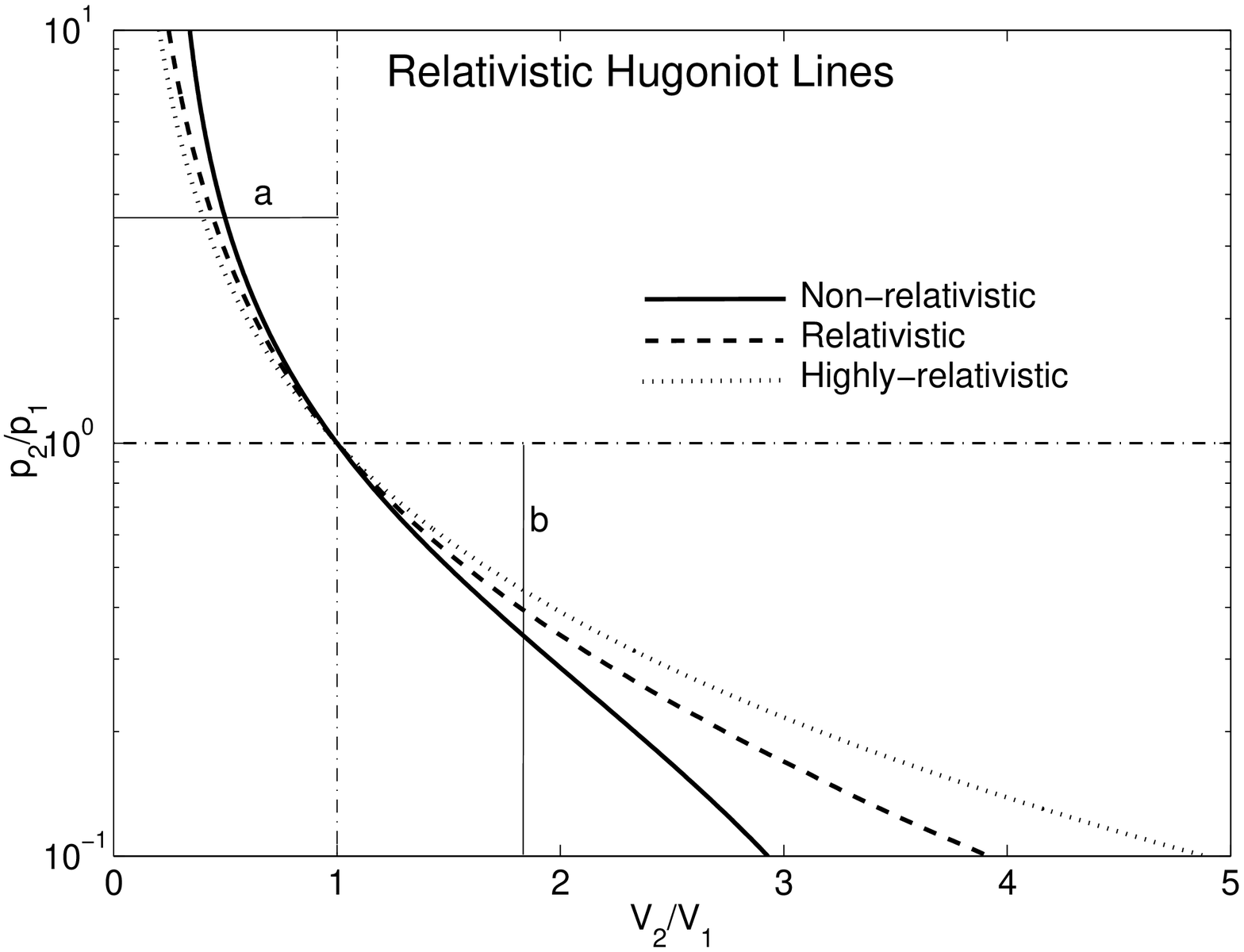}
\caption{ Hugoniot lines for non-relativistic (Equation (\ref{equ:hugoniot non}) with $\Gamma=5/3$), relativistic
  (Equation (\ref{equ:Hugoniot 1}) with $\hat{c}=1$ and $\Gamma=3/2$)
  and highly-relativistic fluids (Equation (\ref{equ:Hugoniot super}) with $\Gamma=4/3$).
The ordinate is the downstream pressure relative to the upstream one; the abscissa is the downstream specific
  volume relative to the upstream value.
All curves pass through the point (1,1).
We notice that for the same downstream to upstream pressure compression ratio (reference line a), the density increase
  is higher for relativistic fluids than for non-relativistic fluids;
  while for the same downstream to upstream density diluent ratio (reference line b), the pressure reduces less
  in relativistic fluids than in non-relativistic fluids.
\label{Fig:hugoniot1}}
\end{figure}

Hugoniot lines represented by Equation (\ref{equ:Hugoniot 1}) are illustrated in Figure \ref{Fig:hugoniot1} for non-relativistic,
  relativistic and highly-relativistic fluids.
Along reference line a, it is seen that
  the density increase is higher for relativistic fluids than for non-relativistic fluids
  for the same pressure enhancement ratio across the compression shock.
And along reference line b the pressure reduction is smaller in relativistic fluids than in non-relativistic fluids
  for the same density dilution ratio across the rarefaction shock.
These differences are due to the fact that pressure assumes a larger proportion of the total energy in relativistic fluids
  when the flow speed gets closer to the speed of light,
  such as the downstream of rarefaction shocks along reference line b.

\begin{figure}
 \epsscale{1.0} \plotone{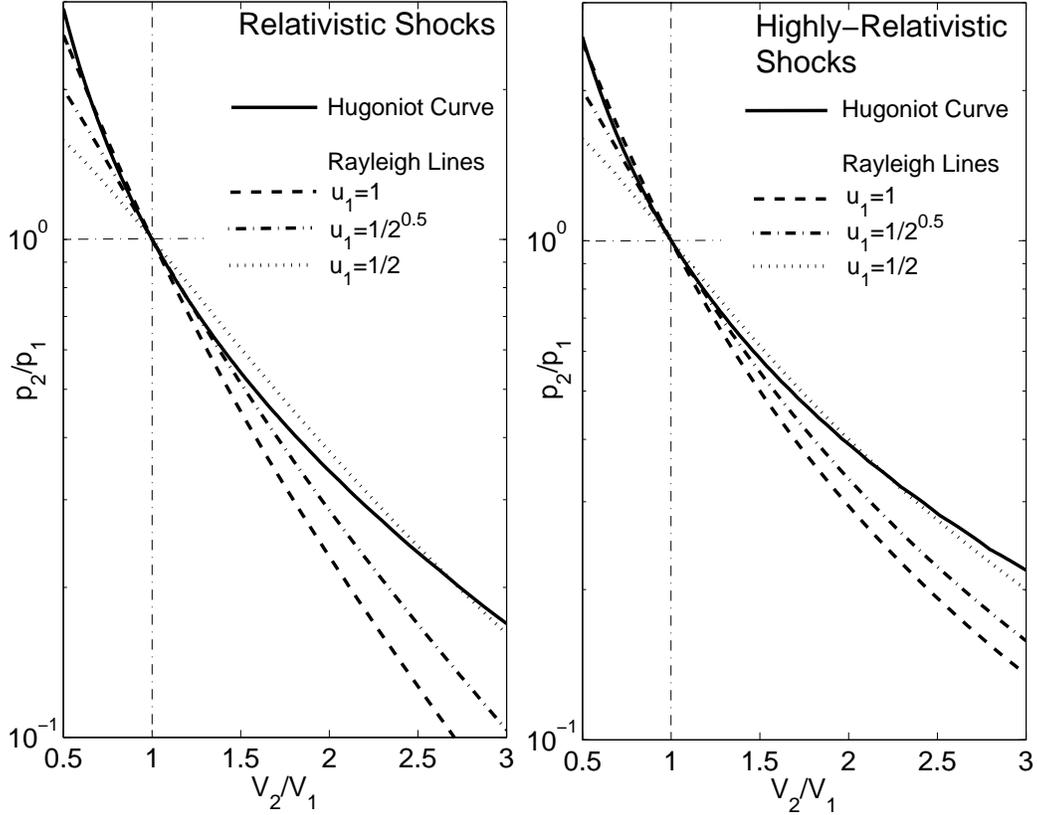}
\caption{ Rankine-Hugoniot shock solutions for relativistic ($\hat{c}=1$, $\Gamma=3/2$) and highly-relativistic
  ($\hat{c}=0$, $\Gamma=4/3$) fluids in the $\hat{p}-\hat{V}$ diagram.
The Rayleigh lines are not straight lines as they are in the non-relativistic case \citep[see, e.g.,][]{landau1959}.
We can see that for (highly-)relativistic fluids, compression shocks with $p_2/p_1>1$ can only be achieved for relatively
  higher upstream fluid speeds (here $u_1=1$ for example).
And for relatively lower upstream fluid speeds (here $u_1=1/2$ for example), only ``rarefaction shocks"
  with $p_2/p_1<1$ (which normally do not happen due to entropy increase) exist.
In between these two kinds of shocks is the sound-speed-upstream-flow, i.e., the Rayleigh line with $u_1=1/\sqrt{2}$,
  which does not have any shock solution.
\label{Fig:shockRH}}
\end{figure}

Typical shock solutions for relativistic and highly-relativistic fluids are shown in Figure \ref{Fig:shockRH}.
One distinct difference from non-relativistic shocks is that the Rayleigh lines for relativistic fluids are not straight lines
  any more, which can be easily seen from the form of Equation (\ref{equ:Rayleigh 1}).
A more essential difference between relativistic and non-relativistic shocks is that compression shocks ($p_2>p_1$, $V_2<V_1$)
  can only be achieved with very high upstream flow speeds in relativistic fluids as indicated for example by the state
  ($u_1=1$, $v_1=\frac{1}{\sqrt{2}}c$) in Figure \ref{Fig:shockRH}.
Furthermore, when the upstream flow speed is reduced by half ($u_1 = 1/2$, $v_1 = \frac{1}{\sqrt{5}}c$),
  only rarefaction shock solution ($p_2<p_1$, $V_2>V_1$) exists.
The criterion distinguishing these two types of shocks is the sound speed of the upstream flow, i.e., $u_1=\frac{1}{\sqrt{2}}$ and
  $v_1=\frac{1}{\sqrt{3}}c$, when the only tangency state between the Rayleigh and Hugoniot lines is the (1,1) point.
This classification of shocks and the criterion are the same as those for the non-relativistic shocks.


\subsection{Negative pressure downstream flows}

\begin{figure}
 \epsscale{1.0} \plotone{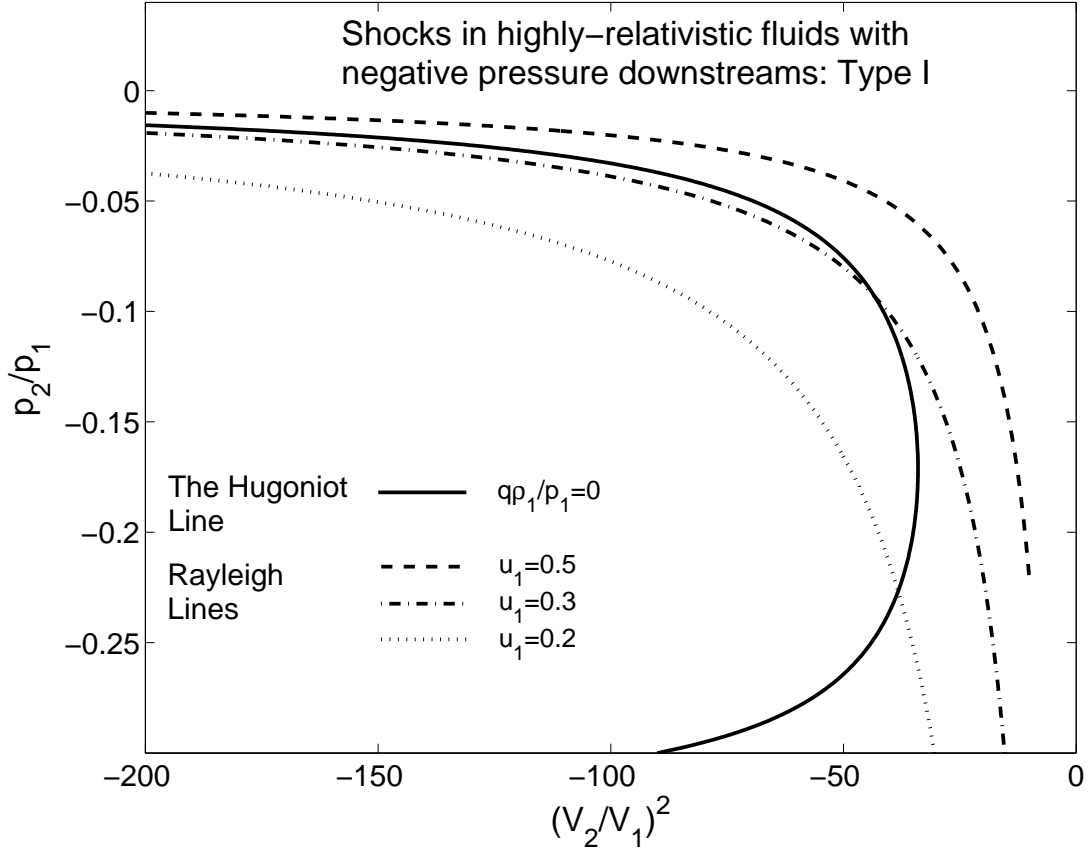}
\caption{ Shock solutions for highly-relativistic ($\hat{c}=0$, $\Gamma=4/3$) fluids in the $\hat{p}-\hat{V}^2$ diagram with negative
  pressure downstream flows.
The square of the specific volume is also a negative value, which means that the downstream specific volume (or density) is
  an imaginary number if we assume the upstream flow to be normal fluids.
We can see that there is no intersection between the Hugoniot line and the Rayleigh line with
  $u_1=0.5$, while intersections exist for Rayleigh lines with $u_1=0.2$ and $u_1=0.3$.
In fact, for highly-relativistic fluids, only when the upstream flow speed is in the range of $0<u_1<\frac{\sqrt{2}}{4}$
  (see Section 2.3) can we achieve negative pressure flows via this kind of shocks.
However all this shocks do not follow the entropy increase law (cf. Section 2.4) so this Type I negative pressure fluids physically
  do not exist.
\label{Fig:shockNegativeP}}
\end{figure}

Under certain conditions of particle flux (\ref{equ:number flux}),
  the downstream pressure can assume negative values.
Take the highly-relativistic case as an example.
Combining equations (\ref{equ:Rayleigh super}) and (\ref{equ:Hugoniot super}), and by setting the adiabatic index to $\Gamma=4/3$,
  we readily obtain the relation between the reduced downstream pressure and the
  upstream four-velocity:
\begin{equation}
u_1^2=\frac{1}{8}(3\hat{p}+1).
\label{equ:negative p}
\end{equation}
Then for upstream velocity in the range $0<u_1<\frac{\sqrt{2}}{4}$, the reduced downstream pressure is negative, i.e.,
  $-\frac{1}{3}<\hat{p}<0$.
Furthermore, under this condition the reduced specific volume
  is an imaginary number according to the highly-relativistic Hugoniot relation (\ref{equ:Hugoniot super}), i.e., $\hat{V}^2<0$.
That is, for shock waves in highly-relativistic fluids with a normal upstream flow ($p_1>0$, $V_1>0$) whose speed $u_1$
  is smaller than $\frac{\sqrt{2}}{4}$, a negative pressure state
  can be achieved in the downstream flow with the specific volume (or density) being an imaginary number.
The $\hat{p}-\hat{V}^2$ diagram showing the highly-relativistic shocks with negative-pressure downstream flows ($u_1=0.2$ and $0.3$)
  is shown in Figure \ref{Fig:shockNegativeP}.
The physical interpretation of this type of negative pressure fluids with imaginary density (defined as Type I) is not clear so far,
  although its existence can be ruled out based on entropy considerations, as will be demonstrated in the sequel.

\begin{figure}
 \epsscale{1.0} \plotone{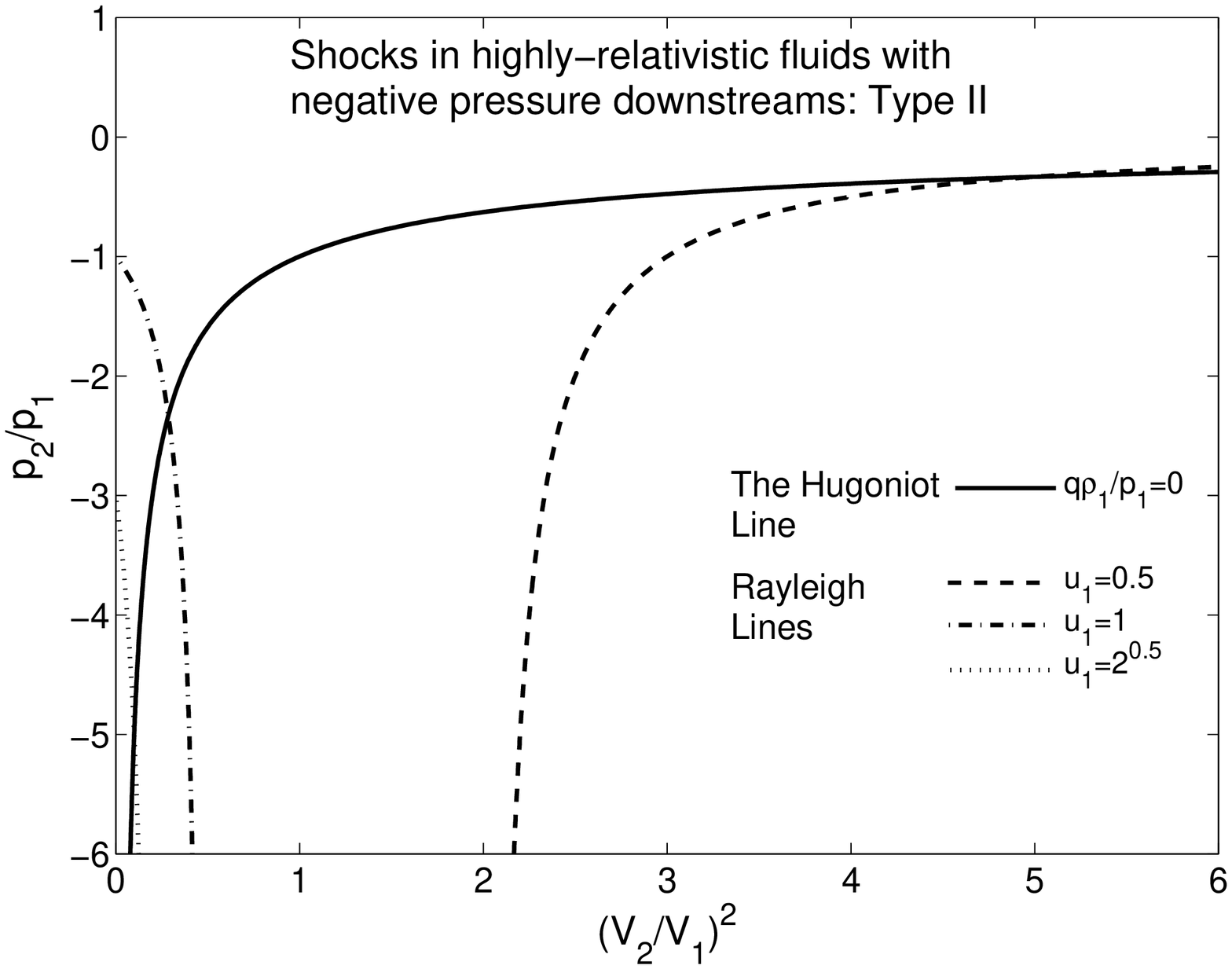}
\caption{Shock solutions for highly-relativistic ($\hat{c}=0$, $\Gamma=2/3$) fluids in the $\hat{p}-\hat{V}^2$ diagram with negative
  pressure downstream flows.
Shock solutions exist for Rayleigh lines with upstream speeds in the range of $u_1>\frac{\sqrt{2}}{4}$, i.e., $u_1=0.5$, $u_1=1.0$
  and $u_1=\sqrt{2}$.
These Type II negative pressure fluids follow the entropy increase law (cf. Section 2.4) thus are physically available.
\label{Fig:shockNegativeP2}}
\end{figure}

Another type of negative pressure fluids (Type II) is the case of $\Gamma=2/3$ in the EoS (\ref{equ:EoS}), i.e.,
  \begin{equation}
  p=-\frac{1}{3}\rho \epsilon,
  \label{equ:Eos dark energy}
  \end{equation}
  which can be interpreted on the basis of dark energy in the universe.
Under this equation of state, the combination of highly-relativistic Rayleigh (\ref{equ:Rayleigh super}) and
  Hugoniot (\ref{equ:Hugoniot super}) relations readily leads the form of the upstream four-velocity:
\begin{equation}
u_1^2=\frac{1}{8}(-3\hat{p}+1).
\label{equ:negative p 2}
\end{equation}
For the upstream flow speed $u_1>\frac{\sqrt{2}}{4}$, the reduced pressure is negative, accounting for a negative pressure
  downstream fluids with the upstream fluids being normal.
This type of negative pressure fluids has a real-number specific volume (or density) according to
  equation (\ref{equ:Hugoniot super}).
Shock waves involving Type II negative pressure fluids are shown in Figure \ref{Fig:shockNegativeP2}.

Entropy analysis showing the availability of these two types of negative pressure fluids are presented in the following.

\subsection{Existence of rarefaction shocks}

All shock waves should follow the law of entropy increase.
Applying the thermodynamic relation \citep{landau1959}
  \begin{equation}
  d\bigg(\frac{\omega}{n}\bigg)=Td\bigg(\frac{\sigma}{n}\bigg)+\bigg(\frac{1}{n}\bigg)dp
  \label{equ:thermal}
  \end{equation}
  in the weak shock wave\footnote{The weak shock wave assumes the discontinuity in every quantity to be small.},
  we have the following form through Taylor expansion:
  \begin{equation}
  \frac{\omega_2}{n_2}-\frac{\omega_1}{n_1}=T_1\bigg(\frac{\sigma_2}{n_2}-\frac{\sigma_1}{n_1}\bigg)+\frac{1}{n_1}(p_2-p_1)
  +\frac{1}{2}\bigg(\frac{\partial (1/n)}{\partial p_1}\bigg)_s (p_2-p_1)^2
  +\frac{1}{6}\bigg(\frac{\partial^2 (1/n)}{\partial p_1^2}\bigg)_s (p_2-p_1)^3.
  \label{equ:thermal taylor}
  \end{equation}
Here $T$ is the temperature, $\sigma$ the entropy per unit proper volume and $s=\sigma/n$ the entropy
  per particle.
From equation (\ref{equ:shock adiabatic}), we have another expression of $\frac{\omega_2}{n_2}-\frac{\omega_1}{n_1}$, i.e.
  \begin{equation}
  \frac{\omega_2}{n_2}-\frac{\omega_1}{n_1}=\frac{1}{2}\bigg(\frac{1}{n_2}+\frac{1}{n_1}\bigg)(p_2-p_1).
  \label{equ:thermal adiabatic}
  \end{equation}
It should be noticed that in achieving equation (\ref{equ:thermal adiabatic}), the higher order term
  $\bigg(\frac{\omega_2}{n_2}-\frac{\omega_1}{n_1}\bigg)(p_2-p_1)$ has been omitted.
Combining equations (\ref{equ:thermal taylor}) and (\ref{equ:thermal adiabatic}), we readily obtain the relation of the entropy
  difference across the shock front:
  \begin{equation}
  T_1\bigg(\frac{\sigma_2}{n_2}-\frac{\sigma_1}{n_1}\bigg)=\frac{1}{2}\bigg(\frac{1}{n_2}-\frac{1}{n_1}\bigg)(p_2-p_1)
  -\frac{1}{2}\bigg(\frac{\partial (1/n)}{\partial p_1}\bigg)_s (p_2-p_1)^2
  -\frac{1}{6}\bigg(\frac{\partial^2 (1/n)}{\partial p_1^2}\bigg)_s (p_2-p_1)^3.
  \label{equ:entropy 1}
  \end{equation}
By introducing the expansion of $\frac{1}{n_2}$ with respect to $p_2-p_1$
  \begin{equation}
  \frac{1}{n_2}-\frac{1}{n_1}= \bigg(\frac{\partial (1/n)}{\partial p_1}\bigg)_s (p_2-p_1)
  +\frac{1}{2}\bigg(\frac{\partial^2 (1/n)}{\partial p_1^2}\bigg)_s (p_2-p_1)^2
  \label{equ:thermal volume}
  \end{equation}
  to equation (\ref{equ:entropy 1}), we have the simplified expression of the entropy difference across a relativistic shock:
  \begin{equation}
  \frac{\sigma_2}{n_2}-\frac{\sigma_1}{n_1}=s_2-s_1=\frac{1}{12T_1}\bigg(\frac{\partial^2 (1/n)}{\partial p_1^2}\bigg)_s (p_2-p_1)^3.
  \label{equ:entropy 2}
  \end{equation}
The law of entropy increase requires that $s_2-s_1>0$.

From the equation of state (\ref{equ:EoS}), the particle number density can be expressed as
  \begin{equation}
  \frac{1}{n}=\frac{(\Gamma-1)m_0\epsilon}{p}.
  \label{equ:number density}
  \end{equation}
By using this expression in the entropy equation (\ref{equ:entropy 2}), we have
  \begin{equation}
  \frac{\sigma_2}{n_2}-\frac{\sigma_1}{n_1}=\frac{(\Gamma-1)m_0\epsilon_1}{6T_1p_1^3}(p_2-p_1),
  \label{equ:entropy 3}
  \end{equation}
  from which it is easily identified that for normal upstream fluids with $p_1>0$ and $\epsilon_1>0$,
  rarefaction waves with $p_2-p_1<0$ only exist for the adiabatic index $\Gamma<1$.
So under the regime of weak shock,
  the Type I negative pressure fluids with $\Gamma=4/3$ does not exist due to the violation of entropy increase law;
  while Type II negative pressure fluids with $\Gamma=2/3$ is expected to be a real physical existence.
However, since it is still questionable as whether the entropy analysis in weak shocks can be generalized to normal strong shocks,
  the existence or non-existence of Type I negative pressure fluids in strong shocks needs to be further studied.

\section{Relativistic detonation and deflagration waves}

When exothermic or endothermic reactions are involved in fluids, detonation or deflagration waves can be excited,
  which renders the dynamics different from those of non-reactive flows, as will be analyzed in the following.

\subsection{Rankine-Hugoniot relations for relativistic combustion waves}

Representing the overall energy release per unit mass by $q$, which is positive and negative for exothermic and endothermic
  reactions respectively,
  the specific enthalpies per unit volume across the reaction front\footnote{The reaction front is normally slim in dimension relative
 to the entire flow under consideration.} are
  \begin{eqnarray}
  \omega_1&=&e_1+p_1+\rho q \quad {\rm and} \nonumber \\
  \omega_2&=&e_2+p_2
  \label{equ:enthalpy 3}
  \end{eqnarray}
  for upstream and downstream flows, respectively.
Following the procedure in deriving Equations (\ref{equ:Rayleigh 1}) and (\ref{equ:Hugoniot 1}), the Rayleigh
  and Hugoniot relations for relativistic reactive flows are given by:
   \begin{equation}
  (\hat{p}-1)= -u_1^2 \bigg[\hat{c}^2(\hat{V}-1)+\frac{\Gamma}{\Gamma-1}(\hat{p}\hat{V^2}-1)-\hat{q}\bigg],
  \label{equ:Rayleigh 2}
  \end{equation}
  \begin{equation}
  \hat{c}^2\bigg[\frac{\Gamma+1}{\Gamma-1}(\hat{p}\hat{V}-1)-(\hat{p}-\hat{V}) -2\hat{q} \bigg]
  = \frac{\Gamma}{\Gamma-1}\hat{p}(1-\hat{V}^2) - \frac{\Gamma}{(\Gamma-1)^2}(\hat{p}^2\hat{V}^2-1)
  +\bigg(\hat{p} + \frac{\Gamma+1}{\Gamma-1}\bigg)\hat{q} + \hat{q}^2,
  \label{equ:Hugoniot 2}
  \end{equation}
where the reduced heat release is $\hat{q}=\frac{\rho_1}{p_1}q$.

\begin{figure}
 \epsscale{1.0} \plotone{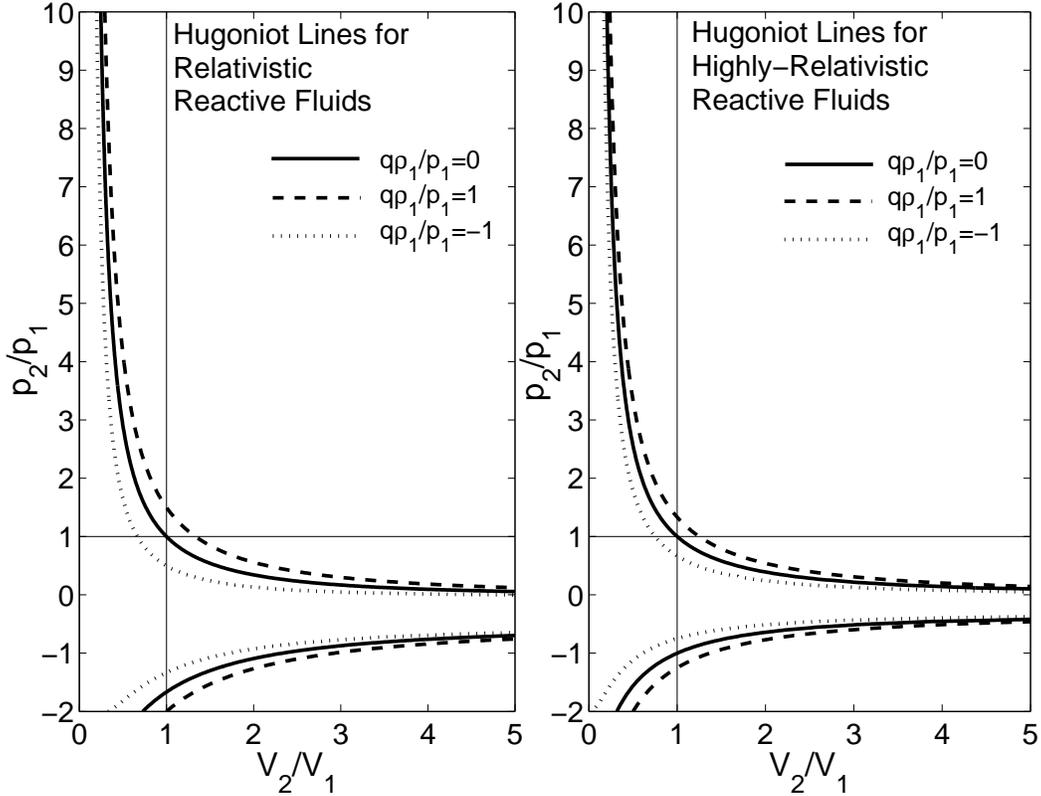}
\caption{ Hugoniot lines for relativistic gas ($\hat{c}=1$, $\Gamma=3/2$)
  and highly-relativistic gas ($\hat{c}=0$, $\Gamma=4/3$).
In the first quadrant,
  the Hugoniot line with positive reaction energy (here $q=1$ for example) is above the shock Hugoniot line with $q=0$,
  and the Hugoniot line with endothermic reaction (here $q=-1$ for example) is below the shock Hugoniot line.
It is also noted that Hugoniot lines are also present in the forth quadrant, hence allowing the existence of shock or
  reactive wave solutions with negative pressures fluids.
One such shock solution has been demonstrated in Figure \ref{Fig:shockNegativeP2}.
\label{Fig:hugoniot2}}
\end{figure}

Representative Hugoniot lines described by relation (\ref{equ:Hugoniot 2}) are shown in
  Figure \ref{Fig:hugoniot2}.
We see that while the non-reactive Hugoniot line passes through the (1,1) point in the $\hat{p}-\hat{V}$ diagram,
  the Hugoniot lines for flows with exothermic and endothermic reactions are above and below the non-reactive one, respectively.
This difference leads to the possible existence of weak detonations in endothermic reactive flows,
  which will be discussed further when considering Figure \ref{Fig:detonationRH2}.


We next note that Rayleigh lines in relativistic reactive fluids
  do not pass through the (1,1) point in the $\hat{p}-\hat{V}$ diagram (see e.g. Figure \ref{Fig:detonationRH1})
  as they do in non-reactive relativistic flows (Figure \ref{Fig:shockRH}).
This is because in relativistic reactive flows, the reaction heat release not only is part of the total energy, but it is also present in the
  momentum conservation equation (see Equations (\ref{equ:momentum cons}) and (\ref{equ:enthalpy 3})),
  which then leads to the presence of the $\hat{q}$ term in the Rayleigh relation (\ref{equ:Rayleigh 2}).

We now separately discuss the solutions for the relativistic detonation and deflagration waves.

\subsection{Detonation waves}

\begin{figure}
 \epsscale{1.0} \plotone{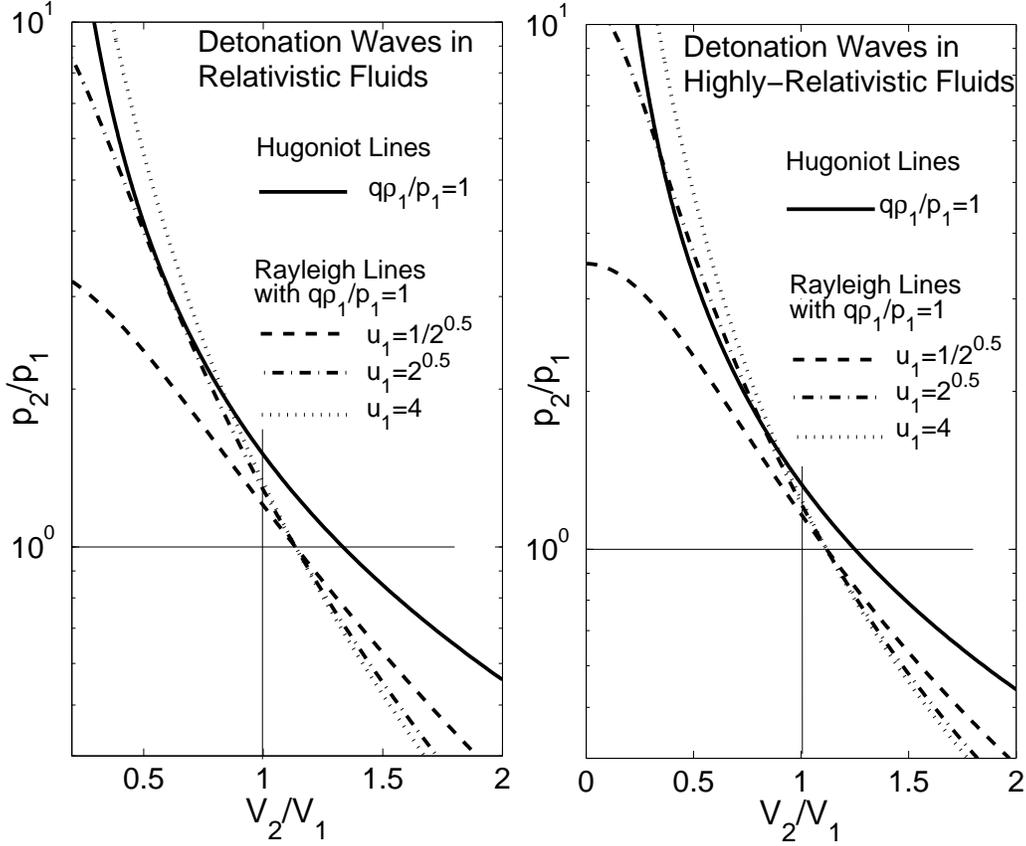}
\caption{ Detonation wave solutions for relativistic gas ($\hat{c}=1$, $\Gamma=3/2$)
  and highly-relativistic gas ($\hat{c}=0$, $\Gamma=4/3$) with an exothermic reaction ($\hat{q}=1$).
For relatively low-speed fluids with $u_1=1/ \sqrt{2}$, there is no intersection between the Rayleigh lines and the Hugoniot lines
  with $\hat{q}=1$, i.e., no exothermic detonation wave exist.
For intermediate-speed fluids with $u_1=\sqrt{2}$, binary intersections can be found between the Rayleigh lines and the Hugoniot
  lines, implying the existence of both strong and weak detonations for the exothermic reactive flows.
For high-speed fluids with $u_1=4$, there are only single intersections between the Rayleigh lines and Hugoniot lines,
  which are weak exothermic detonations.
The criteria for the existence of detonation waves are $u_1=1.0$ and $u_1=0.9$ for the left and right panels, respectively,
  as obtained from numerical explorations.
It is also noticed that for both the left and right panels, all Rayleigh lines intersect at a point with $\hat{q}=1$
  and $\hat{V}>1$.
\label{Fig:detonationRH1}}
\end{figure}

\begin{figure}
 \epsscale{1.0} \plotone{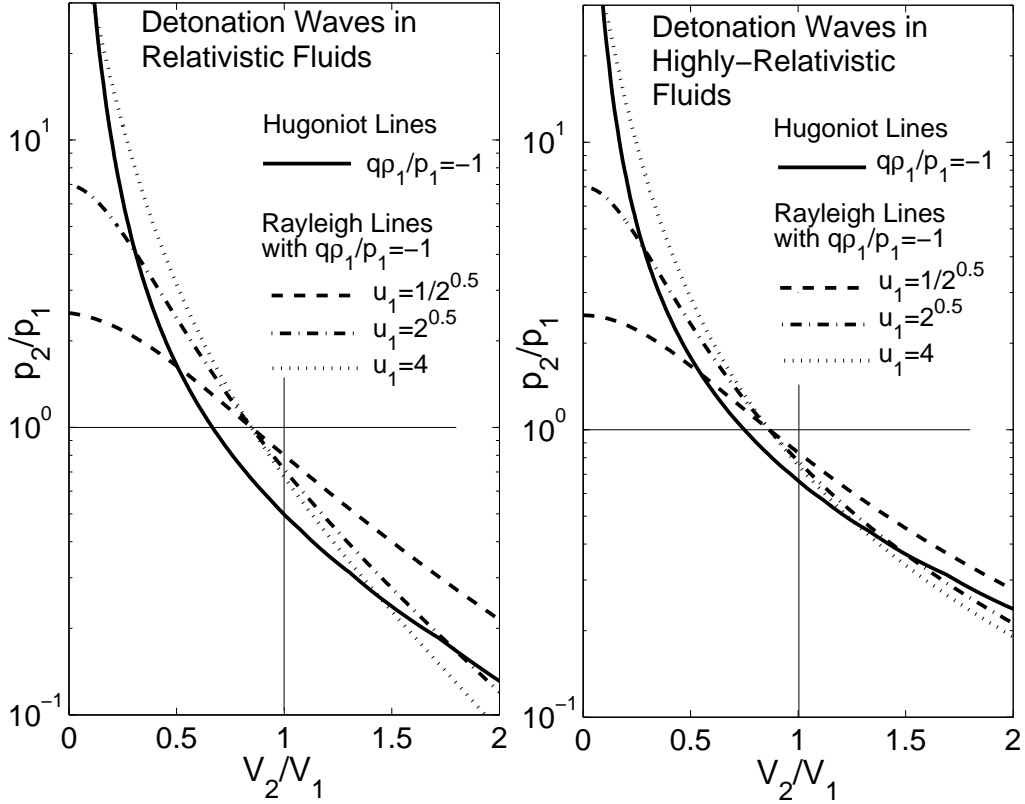}
\caption{ Detonation wave solutions for relativistic gas ($\hat{c}=1$, $\Gamma=3/2$)
  and highly-relativistic gas ($\hat{c}=0$, $\Gamma=4/3$) with an endothermic reaction ($\hat{q}=-1$).
The ordinates are in the logarithm coordinates.
For fluids with all upstream speeds shown here ($u_1=1/ \sqrt{2}$, $u_1=\sqrt{2}$ and $u_1=4$),
  there are single intersections between the Rayleigh lines and Hugoniot lines with $\hat{q}=-1$ in the detonation region ($\hat{p}>1$), which means that weak endothermic detonations always exist.
In both panels, Rayleigh lines intersect at a point with $\hat{q}=1$ and $\hat{V}<1$.
\label{Fig:detonationRH2}}
\end{figure}

Detonation wave solutions for relativistic fluids with exothermic and endothermic reactions are shown in Figure
  \ref{Fig:detonationRH1} and Figure \ref{Fig:detonationRH2}, respectively.

Figure \ref{Fig:detonationRH1} shows that detonation solution exists for exothermic reactive fluids with relatively large upstream speeds,
  i.e., $u_1 \geq \sqrt{2}$; while there is no detonation solution for fluids with relatively slow upstream speed, i.e., $u_1 \leq 1/\sqrt{2}$.
However, the criterion of the upstream flow speed for the existence of detonation is higher than the speed of sound\footnote{In the highly-relativistic fluids, the speed of sound is $u_s=1/\sqrt{2}$; while in normal relativistic fluids, this value is lower.},
  which is different from the case of non-relativistic fluids, for which this criteria is exactly the speed of sound.
This is due to the existence of the heat release $\hat{q}$ in the Rayleigh relation (\ref{equ:Rayleigh 2}) for relativistic
  reactive flows.
By numerically exploring Rayleigh lines with different upstream speed $u_1$, we find that the criteria for the existence of
  detonation wave solutions are $u_1=1.0$ and $u_1=0.9$ for relativistic ($\hat{c}=1$, $\Gamma=3/2$) and highly-relativistic
  ($\hat{c}=0$, $\Gamma=4/3$) gases, respectively, both with the energy release $\hat{q}=1$ (see Figure \ref{Fig:detonationRH1}).
Another interesting feature is that, while flows with $u_1=\sqrt{2}$ have both strong and weak detonation solutions,
  higher speed flows with $u_1=4$ have only weak detonation solutions.
The reason for the disappearance of strong detonation for exothermic fluids with high speed upstream flows is that
  the reaction heat $\hat{q}$ is greatly amplified by the large value of the upstream flow speed
  $u_1$ (this is a relativistic effect), which then leads to large pressure compression ratio $\hat{p}$ to satisfy the energy conservation requirement
  involved in the Hugoniot relation (\ref{equ:Hugoniot 2}).

Detonation solutions in endothermic reactive fluids (Figure \ref{Fig:detonationRH2}) are quite different.
Only weak detonation can be found, and its solution exists for various upstream flow speeds ranging from subsonic
  to supersonic.
The reason for the extensive existence of weak detonations is that endothermic fluids tend to absorb the fluid kinetic energy of any
  strength in order to initiate the endothermic reactions (e.g., ionization of the ISM), which then behave as detonations with
  relatively lower pressure and density compression ratios.
In Figure \ref{Fig:detonationRH2}, the Rayleigh lines for all upstream flows with different speeds have single intersections with
  the Hugoniot line, which shows that there is no threshold for the existence of weak detonations.
On the other hand, strong detonations cannot be formed for flows with any upstream speed.

\subsection{Deflagration waves}

\begin{figure}
 \epsscale{1.0} \plotone{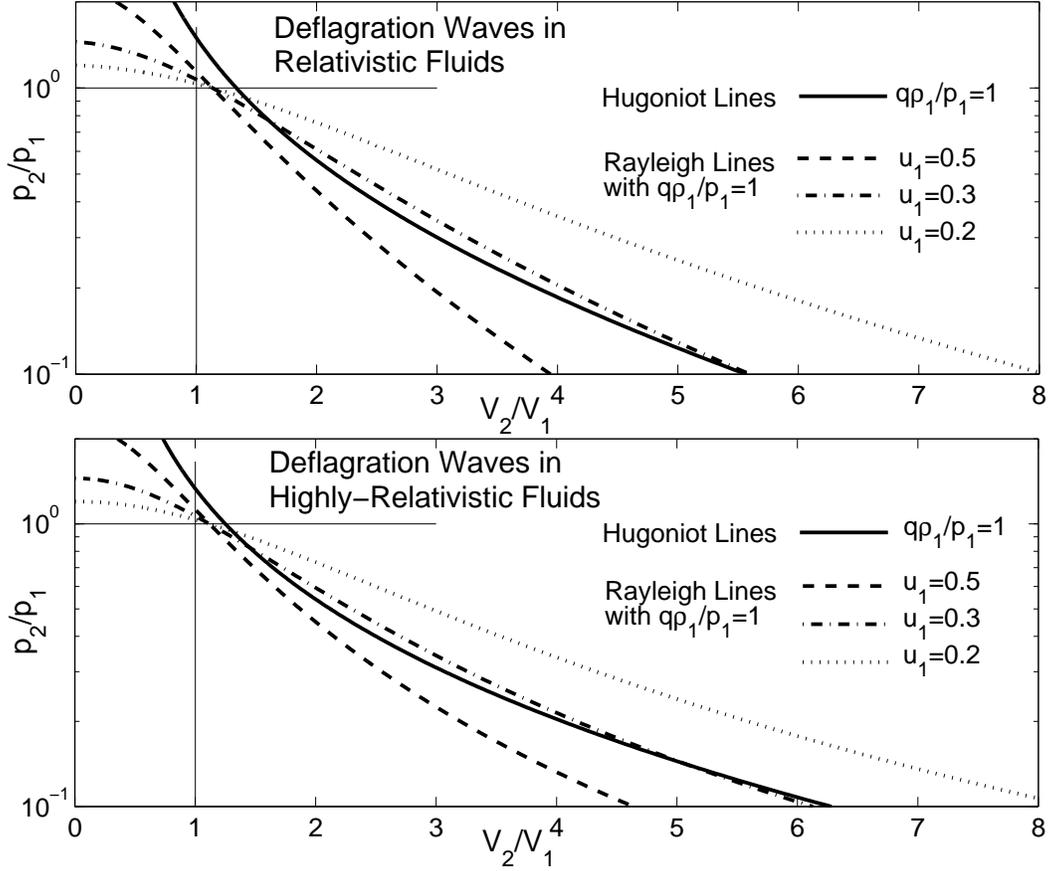}
\caption{ Deflagration wave solutions for relativistic gas ($\hat{c}=1$, $\Gamma=3/2$)
  and highly-relativistic gas ($\hat{c}=0$ and $\Gamma=4/3$) with an exothermic reaction ($\hat{q}=1$).
For relatively high-speed fluids with $u_1=0.5$, there is no intersections between the Rayleigh lines and the Hugoniot lines
  with $\hat{q}=1$, i.e., there is no deflagration waves.
For intermediate-speed fluids with $u_1=0.3$, binary intersections can be found between the Rayleigh lines and Hugoniot lines
  with $\hat{q}=1$, implying the existence of both strong and weak deflagrations for the exothermic reactive flows.
For low-speed fluids with $u_1=0.2$, there are only single intersections between the Rayleigh lines and Hugoniot lines
  with $\hat{q}=1$, representing strong exothermic deflagrations.
The criteria for the existence of deflagrations in the upper and lower panels are $u_1=0.37$ and $u_1=0.39$, respectively,
  as obtained from numerical explorations.
In both panels, Rayleigh lines intersect at a point with $\hat{q}=1$ and $\hat{V}>1$.
\label{Fig:deflagrationRH1}}
\end{figure}

\begin{figure}
 \epsscale{1.0} \plotone{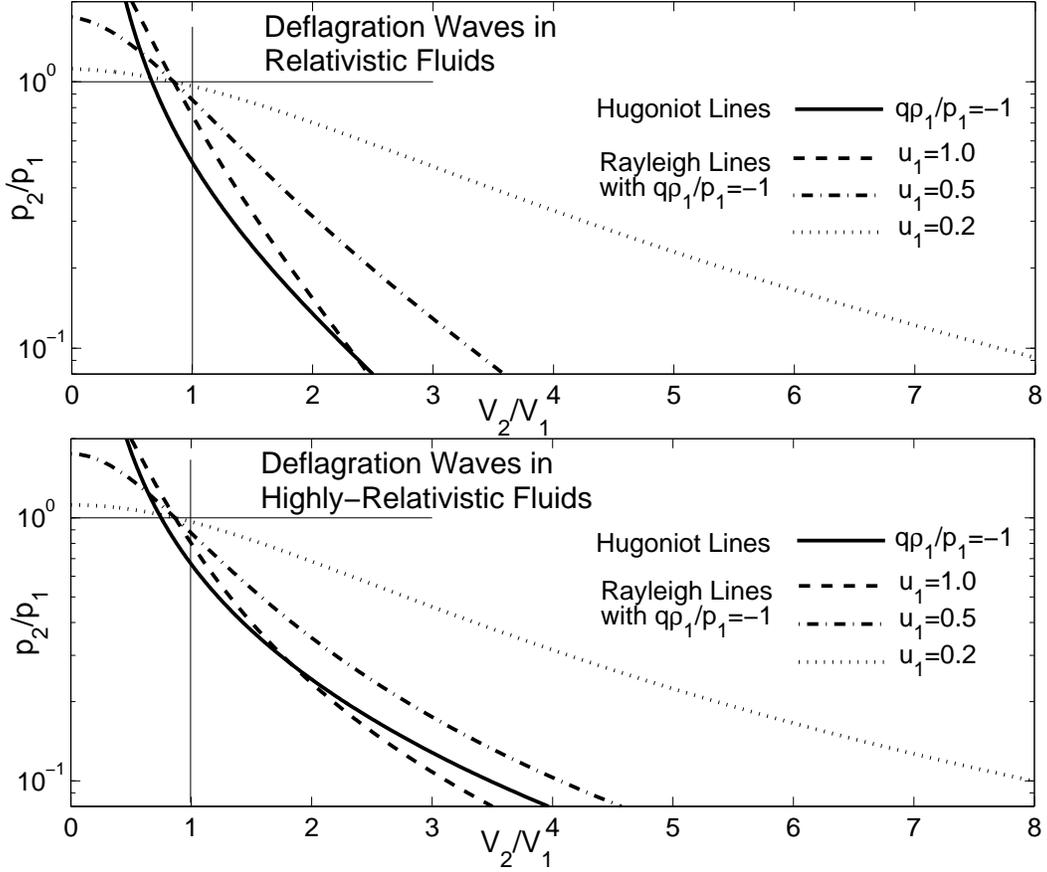}
\caption{ Deflagration wave solutions for relativistic gas ($\hat{c}=1$, $\Gamma=3/2$)
  and highly-relativistic gas ($\hat{c}=0$, $\Gamma=4/3$) with an endothermic reaction ($\hat{q}=-1$).
For low-speed flows with $u_1=0.2$ and $u_1=0.5$, there is no intersection between the Rayleigh lines and Hugoniot
  lines with $\hat{q}=-1$ in the deflagration region ($\hat{p}<1$), which means that endothermic deflagrations do not exist.
For relatively-high-speed flows with $u_1=1.0$, there are intersections between the Rayleigh lines and Hugoniot
  lines with $\hat{q}=-1$ in the deflagration region ($\hat{p}<1$), representing the existence of endothermic deflagrations.
Numerical explorations show that the criteria for the existence of deflagrations are $u_1=0.7$ and $u_1=0.6$ for the upper and lower
  panels, respectively.
In both panels, Rayleigh lines intersect at a point with $\hat{q}=1$ and $\hat{V}<1$.
\label{Fig:deflagrationRH2}}
\end{figure}

Deflagration wave solutions for relativistic fluids with exothermic and endothermic reactions are shown in Figure
  \ref{Fig:deflagrationRH1} and Figure \ref{Fig:deflagrationRH2}, respectively.

Figure \ref{Fig:deflagrationRH1} shows that exothermic reactive fluids have different kinds of deflagration waves for
  different upstream flow speeds.
Specifically, when the upstream flow speed is higher than $u_1=0.5$, there is no deflagration solution in both relativistic
  and highly-relativistic flows.
Taking into consideration that no detonation solution exists for exothermic flows with upstream speeds $u_1\leq 1/\sqrt{2}$
  (see Section 3.2 and Figure \ref{Fig:detonationRH1}), we conclude that for a certain range
  of the upstream flow speed (e.g., $1/2 \leq u_1\leq 1/\sqrt{2}$ in Figure \ref{Fig:detonationRH1} and \ref{Fig:deflagrationRH1}) in relativistic exothermic reactive fluids, neither detonation nor deflagration waves can form.
More specifically, numerical criteria for the existence of deflagration waves have been identified as $u_1=0.37$ and $u_1=0.39$
  for the relativistic and highly-relativistic gases in Figure \ref{Fig:deflagrationRH1}, with the energy release $\hat{q}=1$.
Figure \ref{Fig:deflagrationRH1} further shows that, for the relatively low upstream speed of $u_1=0.3$, both strong and weak
  deflagrations exit;
  but when the upstream speed becomes much lower, i.e., $u_1=0.2$, strong deflagration cannot be achieved, leaving
  only the weak deflagration.
This can be understood by referring to the Rayleigh relation (\ref{equ:Rayleigh 2}), which shows that as the upstream speed $u_1$
  decreases, the pressure rarefaction ratio $\hat{p}$ $(<1)$ should be higher for the same reduced specific volume $\hat{V}$.
Then at a certain low value of $u_1$, there is no intersection between the Rayleigh and Hugoniot lines for higher $\hat{V}$,
  i.e., strong deflagration does not exist.

The wave response is however quite different for endothermic reactive fluids, as shown in Figure \ref{Fig:deflagrationRH2}.
No deflagration solution can be found for flows with upstream speeds of $u_1=0.2$ and $0.5$.
Deflagration waves emerge only when the upstream speed is as high as $u_1=1.0$ (higher than the sound speed).
The numerical criteria for the existence of deflagration waves are $u_1=0.7$ and $u_1=0.6$ for relativistic and highly-relativistic
  gases in Figure \ref{Fig:deflagrationRH2}, respectively, with the energy release $\hat{q}=-1$.
Referring to the deflagration waves with low upstream speed in exothermic reactive fluids, we expect that endothermic deflagrations
  need higher upstream flow speeds to propagate.
Furthermore, even when the upstream flow speed is higher than the speed of sound, shock cannot form because part of
  the fluid kinetic energy is absorbed by the endothermic reaction, which is exactly the case of $u_1=1.0$ in Figure \ref{Fig:deflagrationRH2}.
It is noted (see Section 3.2 and Figure \ref{Fig:detonationRH2}) that the existence of weak detonations for a large range of
  the upstream speed is a distinct feature of endothermic flows.
Consequently weak detonation waves, instead of deflagrations, should be the expected dynamic structure in endothermic
  reactive flows.

\subsection{Chapman-Jouguet waves}

As a parallel analysis to the non-relativistic Chapman-Jouguet waves, the tangency point of the Rayleigh and
  Hugoniot lines, representing a limit case of the wave solutions, is analyzed for relativistic fluids.
At the tangency point the slopes of Rayleigh and Hugoniot lines are equal to each other, so slopes of both curves in the
  $\hat{p}-\hat{v}$ diagram are calculated here according to Equations (\ref{equ:Rayleigh 2}) and (\ref{equ:Hugoniot 2}):
\begin{equation}
\bigg(\frac{d\hat{p}}{d\hat{V}}\bigg)_{\rm Rayleigh} = \frac{\hat{c}^2(\hat{p}-1)+2\frac{\Gamma}{\Gamma-1}\hat{p}\hat{V}(\hat{p}-1)}
  {\hat{c}^2(\hat{V}-1)+\frac{\Gamma}{\Gamma-1}(\hat{V}^2-1)-\hat{q}},
\label{equ:slope R}
\end{equation}
\begin{equation}
\bigg(\frac{d\hat{p}}{d\hat{V}}\bigg)_{\rm Hugoniot} = - \frac{\hat{c}^2(\frac{\Gamma+1}{\Gamma-1}\hat{p}+1)
  +2\frac{\Gamma}{\Gamma-1}\hat{p}\hat{V}+2\frac{\Gamma}{(\Gamma-1)^2}\hat{p}^2\hat{V}}
  {\hat{c}^2(\frac{\Gamma+1}{\Gamma-1}\hat{V}-1)-\frac{\Gamma}{\Gamma-1}(1-\hat{V}^2)+2\frac{\Gamma}{(\Gamma-1)^2}\hat{p}\hat{V}^2-\hat{q}}.
\label{equ:slope H}
\end{equation}
And the solution to the relation
\begin{equation}
\bigg(\frac{d\hat{p}}{d\hat{V}}\bigg)_{\rm Rayleigh}=\bigg(\frac{d\hat{p}}{d\hat{V}}\bigg)_{\rm Hugoniot}
\label{equ:tangency}
\end{equation}
  shows the criterion of the waves represented by the tangency point.
Typical numerical solutions for the Chapman-Jouguet waves as criteria for the existence of detonation waves are given in
  Section 3.2 and Figure \ref{Fig:detonationRH1} captions.

It is noted that in the non-relativistic extreme of $\hat{c}\rightarrow \infty$, the tangency solution degenerates to the normal
  Chapman-Jouguet wave, i.e.,
  \begin{equation}
  \frac{\hat{p}-1}{\hat{V}-1}= - \frac{\frac{\Gamma+1}{\Gamma-1}\hat{p}+1}{\frac{\Gamma+1}{\Gamma-1}\hat{V}-1}.
  \end{equation}
This condition implies that the downstream flow is sonic \citep[c.f.][]{law2006}, which leads to the classical Chapman-Jouguet wave
  in which the downstream flow does not affect the the wave front so that the wave propagates steadily.
However, no simple solution of the tangency point can be found for relativistic fluids as the heat release $\hat{q}$
  are involved in both relations (\ref{equ:slope R}) and (\ref{equ:slope H}).

\subsection{Isobaric waves}

A comparison between Figures \ref{Fig:detonationRH1} - \ref{Fig:deflagrationRH2} and Figure \ref{Fig:shockRH} shows that
  Rayleigh lines for exothermic and endothermic fluids intersect at different points from those of non-reactive fluids in the $\hat{p}-\hat{V}$ diagram.
Specifically, in exothermic reactive fluids, the Rayleigh lines intersect at a point with $\hat{p}=1$ and $\hat{V}>1$;
  while for endothermic reactive fluids, the intersection is at $\hat{p}=1$ and $\hat{V}<1$.
This difference between reactive and non-reactive fluids is caused by the decrease and increase of flow densities in constant-pressure
  exothermic and endothermic waves, respectively.
The example of Figure \ref{Fig:detonationRH2} shows that because the intersection of Rayleigh lines for endothermic reactive fluids
  is to the right of the Hugoniot line, one can easily achieve weak detonation solutions.

Since the intersections of Rayleigh lines for reactive fluids with different reaction heats $\hat{q}$ have the same
  reduced pressure $\hat{p}=1$, we expect the existence of a common solution with $\hat{p}=1$ for any given reaction heat $\hat{q}$.
Then the Rayleigh relation (\ref{equ:Rayleigh 2}) can be readily converted to the following form
  by considering $(\hat{p}-1$), instead of $\hat{p}$, as a common factor:
  \begin{equation}
  (\hat{p}-1)=\frac{-u_1^2\bigg[\frac{\Gamma}{\Gamma-1}\hat{V}^2+\hat{c}^2\hat{V}-(\hat{c}^2+\hat{q}+\frac{\Gamma}{\Gamma-1})\bigg]}
  {1+u_1^2\hat{V}^2\frac{\Gamma}{\Gamma-1}}.
  \label{equ:Rayleigh 3}
  \end{equation}
If we calculate the (positive) $\hat{V}$ root of the numerator in the RHS of this equation, the coordinate of
  the intersection of the Rayleigh line is
  \begin{equation}
  \hat{p}=1, \quad \hat{V}=\frac{\Gamma-1}{2\Gamma}\bigg[-\hat{c}^2+\sqrt{\hat{c}^4+
  4\frac{\Gamma}{\Gamma-1}\bigg(\hat{c}^2+\hat{q}+\frac{\Gamma}{\Gamma-1}\bigg)}\bigg].
  \label{equ:Rayleigh Intersection}
  \end{equation}
In the non-relativistic limit, $\hat{c}\rightarrow \infty$, the intersection is at $(\hat{p}=1,\hat{V}=1)$; while in the
  highly-relativistic limit, $\hat{c}\rightarrow 0$, the intersection is at
  $(\hat{p}=1,\hat{V}=\sqrt{1+\frac{\Gamma-1}{\Gamma}\hat{q}})$.
  (See Figures \ref{Fig:detonationRH1} - \ref{Fig:deflagrationRH2} for the variation of the intersection.)
In the limiting case of small reaction heat in highly-relativistic fluids, i.e.,
  $\hat{q}=\frac{\rho_1}{p_1}q \ll 1$,
 the $\hat{V}$ axis of the intersection can be expressed as $\hat{V}=1+\frac{\Gamma-1}{2\Gamma}\hat{q}$, which can be
  readily converted to the form of
  \begin{equation}
  V_2-V_1=\frac{\Gamma-1}{2\Gamma}\frac{q}{p_1}.
  \label{equ:Rayleigh Vshift}
  \end{equation}
The above equation implies that in an isobaric wave, the specific volume increases and decreases for exothermic
  and endothermic reactions respectively.
In other words, the flow density decreases in an exothermic reaction as the fluid releases energy and increases in an
  endothermic reaction due to the absorption of energy, if the pressure is kept constant.

\subsection{Further considerations of weak detonations and strong deflagrations}

Based on entropy considerations, rarefaction shocks normally do not exist for fluids with the adiabatic index $\Gamma>1$
  \citep[cf. Section 2.4,][]{landau1959,zeldovich1966}, so those weak detonation as well as strong deflagration waves involving rarefaction shocks also normally do not exist.
However, in non-relativistic fluids, weak detonation and strong deflagration waves can be found in systems with
  endothermic reactions or phase transitions, in which a rarefaction shock is not necessary \citep{axford1961,williams1985}.
While these understandings still apply in exothermic relativistic detonations and deflagrations, i.e., the weak detonation
  solutions in Figure \ref{Fig:detonationRH1} and strong deflagration solutions in Figure \ref{Fig:deflagrationRH1}
  should normally not exist in realistic systems,
  there is an exception in that for very high upstream flow speeds (e.g., Rayleigh lines with $u_1=4$ in Figure \ref{Fig:detonationRH1}), weak detonation waves could exist as no rarefaction shock structure is required in such detonations.
Instead, these exothermic weak detonations with high upstream speeds are similar to the weak detonation in endothermic reactive fluids
  (Figure \ref{Fig:detonationRH2}),
  as both reactions can be directly ignited in the shockless waves because the kinetic energies of the upstream flow is large enough
  to initiate the reaction.
It is therefore reasonable to expect that weak detonation as well as strong deflagration waves of this kind are more common in
  high-speed astrophysical flows,
  and they are expected to have different structures from the traditional
  Zel'dovich - von Neumann - D\"{o}ring (ZND) structure in strong detonation waves \citep[cf.][]{williams1985}.
One such example is that of relativistic endothermic reactive fluids, for which weak detonation waves instead of
  weak deflagration waves are the dominant dynamics
  as can be seen from Figure \ref{Fig:detonationRH2} and Figure \ref{Fig:deflagrationRH2}.





We further note that the non-relativistic form of endothermic combustion waves has been studied for interstellar gas ionization
  \citep{axford1961}.
The ionization of neutral hydrogen atoms is an endothermic reaction which forms an `ionization front' in the interstellar media.
This front, separating the unionized (H I) and the fully ionized (H II) regions, is dynamically identical to the detonation
  (or deflagration) front.
\citet{axford1961} has shown the existence of both weak detonation and strong deflagration waves theoretically and what we discussed
  here is a generalization of these results to the relativistic regime.
Besides the ionization of neutral hydrogen atoms around hot stars, the re-ionization (also of neutral hydrogen atoms by stars and
  quasars) process of the universe \citep{miralda2003} can also be considered as an endothermic combustion wave.
Another endothermic fluid process in astrophysics is the inverse Compton scattering, the mechanism for a large variety of high energy
  X-ray and $\gamma$-ray radiations \citep[e.g.,][]{dejager1992,tavani2011}.

For relativistic fluids with $\Gamma<1$, rarefaction shocks are allowed through entropy consideration (cf. Section 2.4),
  and as such lead to the formation of weak detonations.
This is obviously another way of having weak detonations in astrophysical relativistic fluids.

\section{Conclusions}

As a theoretical foundation to study the dynamics of astrophysical systems, the basis of relativistic reactive fluid dynamics
  is formulated here in terms of the Rankine-Hugoniot relations for detonation and deflagration waves in normal relativistic
  reactive flows.
The relativistic shock theory is revisited and the mathematical solutions of two types of negative pressure downstream
  flows are identified.
Normal relativistic and highly-relativistic detonation and deflagration wave solutions are constructed for both exothermic and
  endothermic reactive flows.
The existence of endothermic reactions in astrophysical phenomena extends the family of potentially realizable combustion waves
  in terms of weak
  detonations and strong deflagrations, which do not exist extensively in terrestrial situations.
These theoretical results can be applied in astrophysical systems such as supernovae explosions, $\gamma$-ray bursts and
  the false vacuum decay in the early universe, for which the conventional shock theories do not hold.


\acknowledgments
This work was supported by the start up fund for the
  Center for Combustion Energy at Tsinghua University and the National Science Foundation of China (NSFC 51206088).
YG in addition acknowledges support from the China Postdoctoral Science Foundation No. 2011M500313.

\clearpage



\clearpage

\end{document}